\documentclass[12pt,a4paper]{article}
\usepackage{german,amsmath, amsthm, amssymb}
\usepackage{enumerate}
\usepackage{multicol}
\usepackage{epsfig} 
\usepackage{graphicx}
\usepackage[latin1]{inputenc}
\usepackage[T1]{fontenc} 
\usepackage[english]{babel}
\usepackage{dsfont}
\usepackage{cite} 
\usepackage[font=small]{quoting} 

\newcommand{\R}{\mathbb R}

\theoremstyle{definition}

\title{The ontology of Bohmian mechanics }
  \author{Michael Esfeld$^1$, Dustin Lazarovici$^2$,
          Mario Hubert$^1$, Detlef D\"urr$^2$\\[1.5ex]
$^1$University of Lausanne, Department of Philosophy\\
                 \small Michael-Andreas.Esfeld@unil.ch, Mario.Hubert@unil.ch\\
$^2$Ludwig-Maximilians University Munich, Institute of Mathematics\\
\small   lazarovici@mathematik.uni-muenchen.de, duerr@mathematik.uni-muenchen.de}

\date{}

\begin{document}

\maketitle

\begin{abstract}
\noindent The paper points out that the modern formulation of Bohm's quantum theory known as Bohmian mechanics is committed only to particles' positions and a law of motion. We explain how this view can avoid the open questions that the traditional view faces according to which Bohm's theory is committed to a wave-function that is a physical entity over and above the particles, although it is defined on configuration space instead of three-dimensional space. We then enquire into the status of the law of motion, elaborating on how the main philosophical options to ground a law of motion, namely Humeanism and dispositionalism, can be applied to Bohmian mechanics. In conclusion, we sketch out how these options apply to primitive ontology approaches to quantum mechanics in general.
\end{abstract}


\tableofcontents{}

\section{Introduction}
Bohmian mechanics provides for an ontology of non-relativistic quantum mechanics in terms of particles and their trajectories in physical space and time. It is a mathematically precise quantum theory of particles that grounds the formalism and the predictions of textbook quantum mechanics. The Bohmian law of motion is expressed by two equations, a guiding equation for the configuration of particles in three-dimensional space and the Schr\"odinger equation, describing the time-evolution of the wave-function, which enters the guiding equation. The mathematical form of the law is the following one:

\begin{align*}
\large
\begin{cases} \frac{\mathrm{d} Q(t)}{\mathrm{d}t} &= \; v^{\Psi_t}\bigl(Q(t)\bigr) \hspace*{1cm} \text{\normalsize (guiding equation)}\\
i \hbar \frac{\partial \Psi_t}{\partial t} & = \hspace*{.5cm} H \Psi_t \hspace*{1.8cm} \text{\normalsize (Schr\"odinger's equation)}
\end{cases}
\end{align*}
In these equations, $Q$ denotes the spatial configuration of $N$ particles in three-dimensional space and $\Psi_t$ the wave-function of that particle configuration at time $t$. The guiding equation involves the wave-function whose role is to yield a velocity vector field along which the particles move. The theory is a first order theory; that is to say, given the wave-function it is sufficient to specify the positions of all particles at a given time to calculate future and past motion.\\
When Bohm reintroduced the theory in 1952 (after de Broglie's (1928) presentation of the equations) (Bohm 1952, see also Bohm and Hiley 1993, Holland 1993), he chose to write the wave-function in polar form for which the Schr\"odinger equation splits into two equations. One of them is analogous to the Hamilton-Jacobi equation albeit with an additional term, which he called the quantum potential. He then set out to explain various aspects of the motion of particles in terms of this quantum potential, an explanation that allowed to appeal to Newtonian intuitions. He called the theory a causal interpretation of quantum mechanics, as if the ``quantum force'', the gradient of the quantum potential, could be regarded as the cause of certain strange movements of the particles.\\
The theory, being a first order theory no matter what, is however not Newtonian, and the explanatory value of Newtonian concepts like the one of the quantum potential is dubious. The version of the theory known today as Bohmian mechanics is committed only to particles and their positions; these are the local beables in Bell's sense (Bell 1987, ch. 7) or the primitive ontology in the sense of Goldstein (1998). Furthermore, it is committed to a law of motion that describes how the positions of the particles develop in time. That's all (see notably D\"urr, Goldstein and Zangh\`i 1997, Goldstein and Teufel 2001 and D\"urr and Teufel 2009 for a textbook presentation). In particular, Newtonian concepts like acceleration and forces do not enter the formulation of the theory, and velocities are not an additional, independent parameter.\\
Bohmian mechanics, thus conceived, raises the question of the status of the law of motion and thereby the one of the status of the wave-function, since the wave-function figures in the law. This question has not been answered in a satisfactory manner in the philosophical literature as yet. The aim of this paper is to investigate the possible answers to this question. To start with, we recall the open questions that the traditional view of considering the wave-function as an element of physical reality that guides the motion of the particles faces (section 2). We then discuss two possible stances for grounding the law of motion in the primitive ontology of Bohmian mechanics: the Humean one that regards the law as a contingent regularity (section 3) and the one that anchors the law in a disposition of motion of the particles (section 4). We point out the advantages of dispositionalism (section 5). In conclusion, we briefly explain how the results of this paper apply to primitive ontology approaches to quantum mechanics in general (section 6).

\section{What is the ontological status of the wave-function?}
In order to understand the Bohmian law of motion, it is essential to have a clear idea about the objects appearing in its mathematical formulation. Bohmian mechanics starts from the concept of a universal wave-function, figuring in the fundamental law of motion for all the particles in the universe. That is, $Q(t)$ describes the configuration of all the particles in the universe at time t and $\Psi_t$ is the wave-function at time t, guiding the motion of all particles taken together. To apply the theory in physical situations of interest one uses concepts that allow the practical description of subsystems of the universe. This holds essentially for any fundamental physical theory. In Bohmian mechanics, the appropriate concept is that of an effective wave-function, in terms of which the description of a Bohmian subsystem is again Bohmian, that is, provided by equations of the same form as above. The status of those equations hence is different depending on whether we consider the physical description of the universe as a whole or of a subsystem thereof. Only the law of motion for all the particles in the universe together can be taken as fundamental, as opposed to effective descriptions of subsystems that are (fortunately) of the same form, but not on the same footing.

The effective wave-functions are the Bohmian analogue of the usual wave-functions familiar from the standard formulation of quantum mechanics. They are the formal objects in the theory that are supposed to be epistemically accessible through local experiments, by preparation or by statistical analysis of microscopic systems. It is however important to bear in mind that they are not primitive, but derived from the universal wave-function and the actual spatial configuration of all the particles ignored in the description of the respective subsystem. The non-local law of Bohmian mechanics allows us to encode the influence of those particles, which are not part of the subsystem but nevertheless have an effect on its evolution, in a single object, the effective wave-function, which is defined as a function on the subsystem's configuration space.

This is done in the following way (see D\"urr, Goldstein, Zangh\`i 1992 for more details): if we split the configuration space of the universe into degrees of freedom belonging to the subsystem (denoted by $x$) and the rest of the universe, we may regard only the first as a variable and insert into the universal wave-function the actual configuration $Y(t)$ of all the particles not belonging to the subsystem. The function $\Psi_t(x,Y(t))$ on the configuration space of the x-system then contains all the degrees of freedom necessary to describe its dynamics. In many physically relevant situations it turns out that this function allows the autonomous Bohmian description of the subsystem whence $\Psi_t(x,Y(t))$ becomes the effective wave function $\varphi_t(x)$ figuring in the guiding equation for the subsystem and satisfying an autonomous Schr\"odinger equation. The description of a Bohmian subsystem then requires the specification of both, the positions of the particles constituting the system and the effective wave-function describing their dynamics. The ``state'' of the subsystem at time $t$ is thus naturally represented by a pair $(q_t, \varphi_t)$, where $q_t$ denotes the spatial configuration of the particles and $\varphi_t$ their effective wave-function at time $t$. This ``quantum state'' is no reason for ontological unease, since it supervenes on the primitive ontology and the fundamental law of motion; it is nevertheless an objective, physical degree of freedom belonging to the subsystem (cf. Pusey, Barrett and Rudolph 2012).

How are we to think about those additional degrees of freedom when we consider a Bohmian subsystem? It is clear that they are not -- and cannot be fancied as -- internal degrees of freedom of the particles. Instead, we can achieve an intuitive physical description by conceiving a field or wave deploying in the system's configuration space and determining the possible trajectories for the motion of the particles according to the guiding law. This picture of a pilot-wave or guiding field allows us to make sense of experimental situations such as the interference pattern observed in the double slit experiment: the particle trajectory goes through either one of the slits, but the effective wave-function passes through both and is subject to interference, which manifests itself in the statistical distribution of the arrival points of the particles on the screen after many repetitions of the experiment. If we could not tell this story, the theory would not provide any intuitive physical understanding of the phenomena.

But to what extent shall we take this picture literally? It might seem natural to promote the heuristics to a fundamental ontology when instead of subsystems and effective wave-functions we consider the whole universe and the fundamental universal wave-function. There is indeed a prima facie good reason to do so and to admit the wave-function as a further concrete physical entity in addition to the particles. Bohmian mechanics is not an action-at-a-distance theory. Given the wave-function as a function on configuration space, non-locality is manifested in the fact that the motion of any particle depends on the spatial configuration of the entire system. But the theory, as it stands, does not posit a concrete functional form of the wave-function in terms of the particle positions, and nothing in the formulation of the law suggests a reading of the dynamics in terms of direct particle interactions. Even more, the wave-function is described by a dynamical equation in its own right, an equation that is completely independent of the position or the motion of the particles.

Thus, there are prima facie good reasons to conclude that the wave-function is something in addition to the primitive ontology, if the latter is taken to consist only of particle positions. In admitting the wave-function to the ontology of Bohmian mechanics, it seems, one has identified a physical entity that can account for the particular manner in which the particles move. In other words, acknowledging that Bohmian particles do not per se act on Bohmian particles, it is tempting to maintain that the wave-function does so: the particles move the way they do because the wave-function makes them do so. This seems to be very much the same story as before -- the story of the guiding field --, except that the universal wave-function would have the task to guide the entire configuration of matter in the universe. It is thus no longer the kind of wave that can be prepared or propagate through anything. But first and foremost, the story is now supposed to be the one that nature tells, not just one that physicists tell about experiments. It will thus have to stand up to further scrutiny.

If the wave-function is admitted as part of the ontology, the metaphor of a guiding field may suggest that it has to be understood just like the word says: as a physical field. It is however clear that it cannot be a field (or wave) of the usual kind, existing in three-dimensional physical space. Instead, it would have to be conceived as a field on the high-dimensional configuration space of the universe, implying thus that configuration space is not only a mathematical representation of configurations of $N$ particles in three-dimensional physical space, but has to be granted a physical reality in its own right as a $3N$-dimensional space (see e.g. Bell 1987, p. 128). The ontology then consists of $N$ particles in physical space and the wave-function on configuration space, presumably acting on the one point in configuration space that corresponds, mathematically, to the actual configuration of matter in physical space. But such an ontology is problematic. The mathematical correspondence between points in configuration space and configurations in physical space is obvious. However, if one admits configuration space as a further stage of physical reality -- in addition to and independent of three-dimensional physical space --, it is unclear how there could be a real connection between these two spaces that could amount to something existing in the one space guiding or piloting the motion of entities existing in the other one.

This problem is well-known in the literature (see e.g. Monton 2006 for a clear formulation) and acknowledged even by adherents of Bohm's theory (see e.g. Callender and Weingard 1997, p. 35). There are several proposals that seek to avoid this problem while nevertheless retaining the ontology of a guiding field. One possibility consists in deleting the commitment to configuration space as being part and parcel of the physical ontology and in placing the wave-function in physical space, although it cannot be an ordinary field or wave. Forrest (1988, ch. 6.2) contemplates the idea of the wave-function representing a ``polywave'' in three-dimensional physical space, assigning properties to sets or tuples of $N$ points instead of individual points. Norsen (2010) abandons the commitment to a single wave-function guiding the motion of the particles and associates with each particle an infinite number of fields in three-dimensional space, thus seeking to achieve an ontology that is committed only to local beables. The other, more radical possibility is to take the opposite direction of Forrest and Norsen and to claim that the extremely high-dimensional space known as configuration space is itself the fundamental stage of physical reality, instead of three-dimensional space or four-dimensional space-time. This idea is implemented in Albert's marvellous point formulation of Bohmian mechanics (Albert 1996). In this case, this space is strictly speaking no longer a configuration space, since there is no given configuration of anything to which it is related. In general, however, it is fair to say that none of these attempts has been able to gain widespread acceptance so far (see in particular Monton 2002 and 2006 for a detailed criticism of Albert's proposal).

There is a common argument why the concept of a field (in all its variations) is inappropriate when discussing the nature of the wave-function in Bohmian mechanics. In quantum mechanics, the quantum state of a (closed) system is usually understood projectively. This is to say that it is represented by a ray in a Hilbert-space rather than by a single vector. In Bohmian mechanics (where the position representation is distinguished), this projective nature is manifested in the fact that for any complex number $c$, $\Psi$ and $c\Psi$ give rise to the same equation of motion for the particles (even if we insist on $\Psi$ being normalized, we have the gauge-freedom of multiplying it by any c-number of modulus one). Hence, it can be argued that the particular value that $\Psi$ assigns to individual points in configuration space is meaningless and that the wave-function is therefore not the kind of physical object that we can call a ``field''.

Of course, we should be open-minded enough to concede that the ontology of a physical theory may contain other kinds of objects than just particles and fields. Insisting on an ontological interpretation of the wave-function, the fall-back position may thus be to say that the wave-function is an element of physical reality, although it does not come under any of the familiar categories. But why insist on the ontological interpretation in the first place? In fact, as far as the idea of the wave-function as a concrete physical entity is motivated by the desire to have something in the ontology that can serve as the cause for the motion of the particles, the whole endeavour seems doubtful. Bohmian mechanics does not tell us that particles without a guiding-field are at rest, while particles ``acted upon'' by a wave-function are moving -- or anything of that kind. It rather tells us that certain wave-functions correspond to particles not moving at all, whereas other (generic) wave-functions correspond to particles moving in a particular way, as described by the guiding equation.

In sum, there are well-founded reservations against taking the wave-function to be part of the ontology of Bohmian mechanics (see also Callender unpublished, section 5). The relationship between the wave-function and the motion of the particles is more appropriately conceived as a nomic one, instead of a causal one in terms of one physical entity acting on the other. This leads us to the nomological interpretation of the wave-function, as suggested by D\"urr, Goldstein and Zangh\`i (1997), Goldstein and Teufel (2001) and Goldstein and Zangh\`i (2013).

If one abandons the commitment to the wave-function as a physical entity, the ontology consists only in the particles and their positions, each particle having exactly one, determinate spatio-temporal trajectory. Thus, for instance, in the double slit experiment with one particle at a time, the particle goes through exactly one of the two slits, and that is all there is in the physical world. There is no field or wave that guides the motion of the particle, propagates through both slits and undergoes interference. A non-local law of motion that says that the development of the position of any particle (its velocity and thus its trajectory) depends on the positions of all the other particles, including the particles composing the experimental set-up, accounts for the observed particle positions on the screen. It is precisely for such instances that the non-local influence of the other particles can be encoded in the effective wave-function, but there is no need to commit oneself to the associated intuition of a guiding field as the correct description of physical reality.

A law of motion tells us what happens, or can happen or would happen in four-dimensional space-time (given the specification of initial conditions), but it is not itself an entity existing in space and time. By the same token, the wave-function, insofar as it figures in the law of motion, is a mathematical object defined on configuration space, instead of a physical object existing in addition to the particles. This is to say nothing more than that the formulation of a law of motion for the primitive ontology may contain mathematical objects that do not themselves correspond to physical objects.

The nomological interpretation of the wave-function should obviously not result in replacing the problematic claim that ``the wave-function moves the particles'' with the even more problematic one that ``the law moves the particles''. However, by studying the role of $\Psi$ as it figures in the law, one can learn something informative about the meaning of this thing called ``wave-function''. The answer one arrives at will however depend on what exactly one takes a physical law to be. There are two main views about laws of nature discussed in the philosophical literature. Both of them can be drawn upon for developing the nomological interpretation of the wave-function in Bohmian mechanics. The first possibility is to recognize only particles' positions in the ontology, conceived as categorical properties, and to take the law to supervene on contingent facts about the distribution of the particles in space and time. On this view, the law has a merely descriptive function, but there nevertheless is an objective law of motion formulated in terms of a universal wave-function. The other possibility is to admit more in the ontology than just particles' positions and to take the law, including the universal wave-function, to be grounded in what there thus is added to the ontology. In other words, the law is grounded in the nature or essence of the properties that the entities in space and time instantiate. These properties then are conceived as dispositions (in the sense of what one may call ``law-making properties'', that is, properties for which it is essential to exercize a certain nomological role).

Bohmian mechanics is also compatible with a primitivism about laws as advocated by Maudlin (2007). This view admits the law itself as part of the fundamental ontology. Applied to Bohmian mechanics, this view can limit itself to maintaining that there is nothing more to understand the meaning of the wave-function than to grasp its role in the formulation of the equation of motion for the primitive ontology. Note that according to this view, there is no sharp distinction between a nomological and an ontological interpretation of the wave-function, since laws belong to the stock of physical reality as well. In brief, if one is prepared to accept laws as primitives, there is not much reason to worry about the status of the wave-function in Bohmian mechanics in particular. However, if one does worry about the status of the wave-function, and if one considers Humeanism about laws to be unsatisfactory, then dispositionalism is to our mind the more attractive anti-Humean position (cf. also the review of Maudlin 2007 by Su\'arez 2009, p. 276): dispositionalism offers an account of what the wave-function stands for, instead of merely describing its role in a law that is accepted as primitive. In the following, we set out to show that the ontology of Bohmian mechanics can be clearly stated and understood in terms of a standard metaphysical theory of properties that does not accept laws as primitive, be it a Humean theory, be it a dispositionalist one.

\section{Humeanism about laws}
Insisting on the fact that laws cannot move the entities in space and time, one may draw the conclusion that nothing does so. The view known as Humeanism about laws of nature implements this conclusion, going back to David Hume's denial of necessary connections in nature (e.g. Hume 1748, section VII). According to this view, laws only have a descriptive function. The ontology consists in the distribution of particulars (such as particles and their positions) throughout the whole of space-time. Thus, David Lewis, the most prominent contemporary advocate of this position, advances the thesis of Humean supervenience, characterizing it in the following manner:

\begin{quote} It is the doctrine that all there is to the world is a vast mosaic of local matters of particular fact, just one little thing and then another. (...) We have geometry: a system of external relations of spatio-temporal distance between points. ... And at those points we have local qualities: perfectly natural intrinsic properties which need nothing bigger than a point at which to be instantiated. For short: we have an arrangement of qualities. And that is all. ... All else supervenes on that. (Lewis 1986a, pp. ix-x)\end{quote}

\noindent On this view, the distribution of the fundamental physical properties over the whole of space-time is entirely contingent: for each single token of a fundamental physical property at a space-time point (such as a particle being located at that point), it is conceivable and metaphysically possible to hold that token fixed and to vary all the other tokens. In particular, given a possible world that is a duplicate of the current state of the actual world, the development of the distribution of the fundamental physical properties in that possible world may be entirely different from the development of the distribution of the fundamental physical properties in the actual world. In short, the physical properties instantiated at any given space-time point or region do not impose any restrictions at all on the physical properties that can be instantiated at other space-time points or regions (see Beebee 2006). It is a contingent matter of fact that the distribution of the fundamental physical properties throughout space-time in the actual universe manifests certain regularities. Given that entire distribution, the laws of nature then are, according to Lewis, the axioms of the description of that distribution that achieves the best balance between logical simplicity and empirical content (e.g. Lewis 1973, pp. 72-75).

Humeanism about laws is applicable to Bohmian mechanics. Assume that one knows the positions of all the particles in the universe throughout the whole history of the universe. The wave-function of the universe then is that description of the universe that achieves, at the end of the universe, the best balance between logical simplicity and empirical content. In other words, the wave-function of the universe supervenes on the distribution of the particles' positions throughout the whole of space-time; the same goes for the law of motion. This supervenience relationship includes that the wave-function of the universe applies not only to the actual distribution of particle positions throughout space-time, but also to other possible distributions. Given that we are ignorant about the exact positions of the particles in the universe, we then get, through the quantum equilibrium hypothesis, to effective wave-functions and quantum statistics as the best description we can achieve for subsystems of the universe (cf. D\"urr, Goldstein and Zangh\`i 1992). But note that on this view, only the universal wave-function that supervenes on the particles' positions throughout the whole history of the universe has a nomological status. No effective wave-function describing subsystems can claim such a status.

Since Humeanism is applicable to Bohmian mechanics, one cannot simply jump to the conclusion that the mere fact of quantum entanglement refutes an atomistic worldview such as Lewis' thesis of Humean supervenience (e.g. Teller 1986) or that such a thesis applies only to configuration space by contrast to four-dimensional, physical space-time (Loewer 1996, p. 104). Quantum entanglement is a feature of the formalism of quantum mechanics. The question is what the appropriate ontology for this formalism is. If one elaborates on an ontology of quantum mechanics in terms of particle positions, as does Bohmian mechanics, and if one adopts the metaphysical stance of Humeanism, then one can maintain that (a) the entanglement is a feature only of the wave-function and that (b) the wave-function of the universe supervenes on the distribution of the particle positions throughout the whole of four-dimensional, physical space-time. Consider the following quotation from Bell's paper on ``The theory of local beables'' (1975):

\begin{quote}One of the apparent non-localities of quantum mechanics is the instantaneous, over all space, `collapse of the wave function' on measurement'. But this does not bother us if we do not grant beable status to the wave function. We can regard it simply as a convenient but inessential mathematical device for formulating correlations between experimental procedures and experimental results, i.e., between one set of beables and another. Then its odd behaviour is acceptable as the funny behaviour of the scalar potential of Maxwell's theory in Coulomb gauge. (quoted from Bell 1987, p. 53)\end{quote}

\noindent Bell's remark applies to quantum theories in terms of local beables in general: once one has specified what the local beables are (e.g. particle positions, or what is today known as flashes in Bell's ontology for GRW -- see Bell 1987, ch. 22), one is free to say that the local beables are all there is and that the quantum formalism is a mere means to formulate regularities in the distribution of the local beables in space-time. This is a coherent view (although not necessarily Bell's considered view: when writing about Bohm's theory, he usually urges a commitment to the wave-function as part of the ontology  -- see e.g. Bell 1987, p. 128).

If one takes this view, one is committed only to ``a vast mosaic of local matters of particular fact'' as Lewis puts it in the quotation above, consisting in the case of Bohmian mechanics in particle positions. The mosaic of the particle positions in the actual world happens to be such that on it supervenes an entangled wave-function figuring in a non-local law of motion. But there is no real physical relation of entanglement that exists as a non-supervenient relation in four-dimensional space-time in addition to the relations of spatio-temporal distance among the particle positions. By the same token, there is not any sort of a holistic physical property instantiated in space and time over and above the local particle positions.

There are a number of substantial philosophical objections against 
Humean-ism, which are, however, not relevant to the purpose of this paper (see e.g. Mumford 2004, part I). There is a common objection from physics, which may be taken to be pertinent in our context: on Humeanism, the laws of fundamental physics do not have any explanatory function. They sum up, at the end of the universe, what has happened in the universe; but they do not answer the question why what has happened did in fact happen, given certain initial conditions. In other words, as regards the domain of fundamental physics, Humeanism accepts all there is in that domain as a primitive fact. The reason is the above mentioned possibility of unrestricted combinations: for each single token of a fundamental physical property at a space-time point (such as a particle located at that point), it is conceivable and metaphysically possible to hold that token fixed and to vary all the other tokens. By way of consequence, there are no real connections among the property tokens occurring at space-time points, which could be revealed by a law and which could explain the temporal evolution of the distribution of the fundamental physical property tokens.

However, one can with reason maintain that science in general -- and fundamental physics in particular -- searches for real connections in nature and that the stress that science lays on discovering laws derives from the idea that laws reveal the real connections that there are in nature instead of being mere devices of economical bookkeeping. Let us therefore enquire into another option to ground the laws of nature in the physical properties, namely dispositionalism and its applicability to Bohmian mechanics.

\section{Laws grounded in dispositions}
The main anti-Humean position about laws of nature in contemporary philosophy traces laws back to properties. According to this view, it is essential for a property to induce a certain behaviour of the objects that instantiate the property in question; the law then expresses that behaviour. A stock example is gravitational mass in Newtonian mechanics: in virtue of being massive, particles attract (accelerate) each other in the manner described by the Newtonian law of gravity. Obviously, the ambition of this account is not to tell us something new about the physics of Newtonian gravity, but to ground the law in the ontology, that is, to clarify how theoretical terms (expressed in the language of mathematics) connect to the entities existing in the physical world. The parameter we call ``mass'' refers to a property of the particles. This property is not a pure quality (for then the question how this property connects to a law would remain unanswered), but a disposition whose manifestation is the mutual attraction of the particles as expressed qualitatively and quantitatively by the law.

In general, this view amounts to considering properties as dispositions to bring about certain effects; the laws supervene on the dispositions in the sense that they express what objects can do in virtue of having certain properties (e.g. Dorato 2005, Bird 2007). Consequently, laws are suitable to figure in explanations answering why-questions. On this view, as on Humeanism, laws do not belong to the ontology. However, in contrast to Humeanism, they are anchored in the essence of the properties of the objects that there are in the physical world, instead of being mere means of economical description.

When applying this view to Bohmian mechanics, one has to be aware of the fact that it has nothing to do with the claim that all properties apart from position are contextual in Bohm's theory; that claim is sometimes formulated in terms of all the other properties being dispositions that are actualized in certain measurement contexts. As regards the philosophical literature on properties as dispositions, it is misleading to talk in terms of dispositions in this respect, because dispositions are real and actual properties that exist in the world independently of any measurements. The so-called contextual properties, by contrast, simply are ways in which the particles move in certain contexts such as certain experimental set-ups and statistical descriptions that apply to those contexts (and that can be expressed by self-adjoint operators as bookkeeping devices). But these ``contextual properties'' are not fundamental. Strictly speaking, they are not properties of anything at all, and there is nothing of philosophical interest in them (as there is nothing of philosophical interest in self-adjoint operators). The point at issue is whether the fundamental physical properties in Bohmian mechanics are dispositions and whether it is in this manner that they ground the law of motion.

In a recent paper, Belot (2012) discusses the option of tracing the Bohmian law of motion back to dispositions. Over and above the position of each particle, Belot countenances a disposition that determines the velocity of the particles as an additional, holistic property of the system of particles under consideration. He characterizes this view in the following manner:

\begin{quote}That is, let us explore an interpretation of Bohmian mechanics under which the complete history of a quantum system is specified by specifying for each time t a configuration $q(t$) of the particles together with the dispositional property $\Phi_t$ that tells us, for each possible configuration of the system of particles at $t$, what the velocity of each particle would be were that configuration actual. On this interpretation, all we have are particles and properties of (systems of) particles -- at each time, each particle has a mass and a position and the system of particles as a whole has a further dispositional property. Let us call this approach the \emph{dispositionalist interpretation} of Bohmian mechanics. (p. 78)\end{quote}

\noindent On this view, the universal wave-function $\Psi_t$ of the system of particles at a given time is a mathematical object that represents the disposition to move in a certain manner at that time. This disposition is a holistic property of all the particles in the universe together -- that is, a relational property that takes all the particles as relata. It induces a certain temporal development of the particle configuration, that development being its manifestation. In other words, given a spatial configuration of the particles (actual or counterfactual) and the disposition of motion at a time as represented by the wave-function as input, the Bohmian law of motion yields the velocities of the particles at that time as output.

Again, what needs to be emphasized and may have remained unclear in Belot's exposition, is that only the universal wave-function has a nomological status. It would be a misunderstanding to apply nomological interpretations to wave-functions of any odd physical system and to seek to ground such wave-functions in dispositions. The effective wave-function of a subsystem defines an equation of motion for the particles constituting that subsystem, but it cannot be seen as expressing the disposition of motion of those particles, since only the disposition of motion of the totality of particles in the universe can serve as the ontological basis for the law of motion.
Given that the disposition of motion is a property of all the particles in the universe together, it cannot be but a disposition that produces its manifestation spontaneously, that is, a certain form of motion of the particles; this disposition cannot require an external stimulus, since there is nothing external to the totality of all the particles in the universe. There is no metaphysical reason to hold that dispositions necessarily depend on external triggering conditions for their manifestation. If the fundamental physical properties are regarded as dispositions, it is reasonable to assume that they manifest themselves spontaneously. Coming back to the mentioned stock example of a fundamental physical disposition, in Newtonian gravity, particles spontaneously attract each other in virtue of their mass.

Belot (2012) formulates an objection to the dispositionalist interpretation of Bohmian mechanics, suggesting that dispositions alone may not tell the entire story about the physical matter of facts. Concretely, he considers a single Bohmian particle in a one-dimensional box and points out that there are countably many real-valued wave-functions corresponding to different energy eigenstates, but representing the same disposition of motion, as they all describe a particle that is at rest at all times. He concludes:

\begin{quote} Observations of this sort may seem to all but scupper the dispositionalist interpretation: for it may well appear that in taking dispositional histories rather than histories of the quantum state to be fundamental, the dispositionalist interpretation discards a great deal of the essential physical content of quantum mechanics. (p. 80) \end{quote} 

\noindent Such worries do not apply to the account presented here. We have already emphasized that only the common disposition of motion of all the particles in the universe together is fundamental. But if the ``dispositional histories'' involve the spatial configuration and the disposition of motion for all there is in the universe, there is no physical content left that would be unaccounted for. Note that energy is not a primitive quantity in Bohmian mechanics. The universal wave-function can be an eigenstate of the universal Hamiltonian (and will be, in fact, if it is stationary), but the corresponding eigenvalue has no physical meaning. Energy plays a physical role in the description of subsystems. Hence, we should indeed worry about the energy of the system ``particle in a box'' if we could stick our hand in it. If, however, that single particle was all there is in the entire universe, expressing its law of motion in terms of different real-valued wave-functions, corresponding to different eigenvalues of the Hamiltonian, would not describe different physical facts.

Still, the observation that even the entire history of dispositions of motion does not necessarily determine a unique (universal) wave-function is obviously correct. Many different wave-functions may give rise to the same law of motion. One may find this fact unsatisfactory. However, we have left the idea of treating the wave-function as an additional physical object behind in section 2. If one regards the wave-function as a formal object appearing in the mathematical formulation of the law, there is no need to insist on the relationship between the fundamental disposition of motion and its representation in terms of a universal wave-function on configuration space being one-to-one. Let us bear in mind that mathematical representations need to be unambiguous only as far as the elements of physical reality and the fundamental laws are concerned. It is a further advantage of primitive ontology theories that they enable us to make such distinctions in a clear-cut way.

In a manuscript available on his website, Thomson-Jones (2012) also suggests a dispositionalist reading of the wave-function in Bohmian mechanics. By contrast to Belot (2012) and to our view, Thomson-Jones attributes to each particle a multitude of dispositions for various motions in three-dimensional space whose manifestation depends on the positions of all the other particles (the dispositions of the other particles are the triggering conditions for the dispositions of the particle in question). However, let us recall that the wave-function enters Bohmian mechanics through the role it plays in the law of motion (the guiding equation) by fixing the velocity of the particles and that only the universal wave-function has a nomological status. The rationale for introducing dispositions in the ontology of Bohmian mechanics is to ground the law of motion in the ontology while being committed only to entities that exist in physical space and time. Consequently, since there is only one universal wave-function, and since the universal wave-function is in any case non-separable, there is no reason to commit oneself to anything more than sparse dispositions: there is exactly one disposition that fixes the form of motion of all the particles by fixing their velocities, thus fixing the temporal development of the configuration of particles. That single disposition is sufficient to account for the motion of any particle (or any sub-collection of particles) in all possible circumstances.

There is another common objection to the nomological interpretation in general and its dispositionalist variant in particular, which derives from the following observation: if the fundamental wave-function of the universe is conceived merely as a formal object figuring in the law of motion for the particles, it may seem strange that it follows a dynamical equation itself and hence gives rise to a law of motion which is time-dependent (or rather, ``time-indexed''). In reply to this objection, let us note in the first place that dispositionalism can in principle accommodate the idea of laws being time-dependent. Laws can be grounded in the dispositions that the objects in the universe instantiate at a given time. Thus, if the disposition for a certain form of motion of the configuration of particles in the universe is time-dependent, so is the universal wave-function that is grounded in that disposition and, consequently, the law of motion. This would mean that time is not only parameterizing the motion of the particles, but that the particles are somehow sensitive to the precise value of that parameter, as if they carried a little clock with them. In other words, the particles have not just the disposition to move in a certain way, given a certain spatial configuration, but that disposition also depends on what time it is in the universe. We would find this somewhat strange, but we see no a priori reason why the universe could not be like that.

Nonetheless, dispositionalism about laws, as applied to Bohmian mechanics, is by no means committed to this view. It is reasonable to suppose that if there were two identical spatial configurations of the universe but at different times, these configurations would behave in exactly the same way. In brief: same spatial configuration of all the particles in the universe, same disposition of motion. Taking this view then commits us to the supposition that the universal wave-function that enters the Bohmian guiding equation is time-independent -- a supposition that is in fact less bold than often alleged.

Note that, pace Belot (2012, pp. 74-77), one is not automatically committed to enter into the even more controversial issue of a quantum theory of gravity if one envisages a time-independent Bohmian law of motion: one can conceive a quantum theory with a universal wave-function that is a stationary solution of the Schr\"odinger equation without that quantum theory having to include gravitational degrees of freedom. In Bohmian mechanics, a wave-function can define a non-trivial dynamics for the motion of particles without itself evolving in time. Bohmian mechanics is not hit by the notorious problem of time that plagues the usual approaches to a universal wave-function (especially in quantum gravity). In other words, the common dilemma of either having to countenance a time-dependent wave-function or facing the problem of time does not apply to Bohmian mechanics. The reason is that the wave-function does not provide the ontology of the theory. The ontology, the stuff that moves in space and time (or the geometry of space-time itself), is external to the wave-function. The latter only enters into the equation of motion for the entities posed in the primitive ontology. Hence, if one replaces the time-dependent wave-function in the guiding equation with a time-independent one, that equation still yields as output the velocities of the particles in the universe at time $t$. It is only that in this case, it is irrelevant what time it is in order to determine the velocities of the particles, given their positions.

Furthermore, let us bear in mind that it is very well possible for effective wave-functions to have a non-trivial (Schr\"odinger-type) evolution in time, even if the fundamental wave-function of the universe from which they are derived is stationary. The time-dependence of the effective wave-functions then comes only from the evolution of the spatial configuration $Y(t)$ of the particles that are not part of the described subsystem (see Goldstein and Teufel 2001 for details). In brief, the universal wave-function entering the guiding equation may very well be a stationary solution of the universal Schr\"odinger equation and still account for the usual quantum behaviour of subsystems as described by effective wave-functions and their Schr\"odinger equations. The Schr\"odinger equation for the universal wave-function is then to be regarded not as a dynamical equation, but (similar to the Wheeler-deWitt equation) as a constraint on the universal wave-function, stipulating its time-independence.

\section{Advantages of dispositionalism}

In section 2, we have pointed out problematic aspects of admitting the wave-function as part of the ontology. These aspects can be divided into the following two ones: the status of configuration space and the ability of the wave-function to guide the motion of particles in physical space. Let us now consider how dispositionalism can improve on these issues.

The spatial configuration of a universe of $N$ particles at time $t$ is represented by a vector $Q(t) \in \R^{3N}$. This vector can be regarded as a point in configuration space, but represents the actual configuration of the particles in three-dimensional physical space. The 3N-dimensional configuration space is usually understood as the mathematical representation of all possible configurations of $N$ particles existing in three-dimensional physical space. Regarding the universal wave-function as a complex-valued function $\Psi$(x), the variable $x$ then ranges over all possible spatial configurations of the universe. A tension now arises when this function is supposed to represent a physical field, existing as a concrete entity in the actual physical world. Then, one either has to argue that the space of possible configurations somehow supervenes on the actual configuration of the particles in physical space, or grant it an independent reality in addition to three-dimensional physical space.

The dispositionalist, by contrast, has no such troubles. In being located in a certain manner in three-dimensional physical space, the particles have the disposition to move in a certain way. The universal wave-function, as a function on configuration space, represents that disposition, concerning, notably, all possible configurations of the particles, actual and counterfactual. At time t, the disposition of motion is instantiated only by the actual configuration of the universe at that time, corresponding, formally, to evaluating the function $\Psi_t$ (or rather the guiding equation) at $Q(t)$. Nonetheless, qua being a disposition and thus a modal property, it grounds the truth value for counterfactual propositions stating how the particles would move if another configuration were the actual one. In brief, on dispositionalism, the truth value of all counterfactual propositions is grounded in what actually exists in the physical world, namely the disposition of motion of the particles.

Concerning the second issue, the commitment to dispositions obviously fulfils the task of providing in the ontology for something that is able to account for the motion of the particles. From a philosophical point of view, nothing speaks against considering the disposition for a certain form of motion as a causal property, although it is not a force or a potential, simply because its essence is to do something, namely to fix a certain velocity with which the particles move. Consequently, the law of motion grounded in this disposition can be regarded as a causal law, giving rise to causal explanations, whereby the efficient cause for the motion of the particles is situated in the particles themselves.

In short, there is a disposition for a certain form of motion as a property of the particles instead of a field or a pilot-wave external to the particles that is supposed to move them. Nonetheless, it may seem that the dualism of entities has simply been replaced with a dualism of properties. On the one hand, one may argue, there is position as an intrinsic and categorical property of each of the particles, and on the other hand, there is a disposition of motion as a relational and dispositional property of the particles. One may also consider the velocity of the particles as a further property.

In reply to this query, let us note the following three points: (1) There is no problem of connection of the various properties admitted in the ontology, if all elements of the ontology are located in physical space and, moreover, if all properties are properties of the particles. In other words, even if there is a dualism of properties, it does not give rise to the objections that hit the dualistic ontology of particles in physical space and a field on configuration space. (2) Insofar as there is a dualism of a disposition for a certain form of motion and the velocity of the particles as its manifestation, it is not troublesome. It rather is an analytic dualism: if one recognizes dispositions, one also has to admit properties that are the manifestations of the dispositions in question (otherwise there would be no dispositions). (3) As regards the relationship between the position of the particles and their disposition for a certain form of motion, one can maintain that position also is a relational property, consisting in a relation of spatial or spatio-temporal distance among the particles (if one does not presuppose the existence of an absolute space or at least does not conceive the location of a particle at a point in absolute space as an intrinsic property of the particle).

Going one step further, the dispositionalist does not even have to concede that position, conceived as spatial or spatio-temporal relations among the particles, is a categorical property by contrast to a dispositional one (so that in the end, there is no dualism of dispositional and categorical properties). For instance, in general relativity theory, one can regard the spatio-temporal, metrical relations as including the disposition to move the particles (since the metrical relations yield the manifestation of gravitation). The Bohmian law of motion, grounded in dispositions, may justify a similar claim: in virtue of standing in certain spatial or spatio-temporal relations, the particles have the disposition to move in a certain manner. It may therefore well turn out that in the end, there is only one type of physical properties, namely relations that are dispositions or that bestow dispositions to move on the system of particles. In any case, one can say that in standing in spatial relations (being localized), the particles have the disposition to move in a certain manner (change in their spatial relations), making thus clear that there is no separation between these two properties.

Let us finally compare dispositionalism with Humeanism. Whereas the Humean admits only particle positions conceived as categorical properties as ontologically primitive, the dispositionalist regards the particle positions as including a disposition for a certain form of motion. Due to this disposition, there are real connections in nature in a twofold sense: first of all, since the disposition of motion is a holistic property of all the particles taken together, it establishes a real connection among the particles -- a real, irreducible relation over and above the external, geometrical relations of spatial distance. Furthermore, since this disposition induces a certain form of motion of the particles, it establishes a real connection (a causal connection having the ontological status of a real connection) between the configuration of the particles in the universe at a given time and that configuration at future times. Basing oneself on this disposition, one can therefore explain the temporal development of the configuration of the particles in the universe, given an initial state (that is, an initial configuration of particle positions).

One may grant that dispositionalism, by contrast to Humeanism, is able to explain the development of the universe by admitting real connections. However, one may object that dispositionalism does not fare better than Humeanism when it comes to explaining the most striking feature of quantum mechanics, namely the non-locality that is, for instance, revealed in the EPR-Bohm experiment (but also already implicit in the double slit experiment). Dispositionalism accepts a disposition of motion as a holistic property of the totality of the particles in the universe as primitive. It can thus not explain why this disposition is a holistic property rather than a property that belongs to each particle considered individually.

However, one can retort that calling for such an explanation expresses a wrong-headed presupposition rather than a sound demand. Quantum non-locality can with reason be taken to be a fundamental feature of nature. Calling for an explanation of it is possible only against the background of assuming that the natural state of affairs is one of locality. Whereas the experience of our macroscopic environment may suggest such a view, it is by no means self-evident; Bohmian mechanics can explain why the non-locality of quantum mechanics is not manifest in everyday experience. But a priori, there is no reason why the temporal development of what there is in the universe should be determined by local rather than by holistic properties.

\section{Conclusion}

In conclusion, let us put the argument of this paper into the wider framework of approaches that consider the formalism of quantum mechanics as representing or referring to a physical reality that is external to this formalism and that cannot be derived from it (by contrast to approaches that accept the quantum formalism, that is the wave function and / or the observables, as irreducible and seek to establish solely on this basis a link with experience). Theories in this vein describe the quantum world in terms of what is known as a primitive ontology -- ``primitive'' because the ontology cannot be read off from the formalism but has to be posed as that to what the formalism refers. In other words, none of the objects appearing in the formulation of textbook quantum mechanics can be straightforwardly taken to represent stuff existing in time and space, but we can achieve an understanding of those objects by inquiring into their relation to the primitive ontology, once the latter has been established as that what the formalism is actually about. Bohmian mechanics is the most elaborate approach in this sense: particles in three-dimensional space make up the primitive ontology, and a guiding equation is provided that describes their motion and into which the quantum mechanical wave-function enters. Other proposals for a primitive ontology include density of stuff in three-dimensional space as in the GRW mass density interpretation (Ghirardi et al. 1995) and flashes occurring at points in four-dimensional space-time as in the GRW flash interpretation (Bell 1987, p. 205, Tumulka 2006) (see Allori et al. 2008 for an illuminating comparison between these approaches).
Primitive ontology approaches face the difficulty that what is posed as the primitive ontology in three-dimensional space does in itself not contain anything that accounts for the temporal development of the elements of that ontology (particle positions in Bohmian mechanics). This shortcoming then motivates an ontological dualism, admitting both the elements of the primitive ontology in three-dimensional space and the wave-function in configuration space as constituting physical reality. However, as pointed out in section 2, it is doubtful whether the wave-function in configuration space is suitable to fill this lacuna. Following the spirit of the idea to regard the formalism of quantum mechanics as a means of representation that refers to a physical reality external to it, one gets to the view of the wave-function being nomological rather than being an element of physical reality -- or, more generally speaking, to the idea of the formalism of quantum mechanics providing for a law, such as the Bohmian law of motion (or the GRW law if the standard formalism is modified to a non-linear equation for the wave-function). It would then obviously be misguided to call for a one-to-one correspondence between the mathematical objects entering the formulation of the law and the elements of physical reality.
The purpose of this paper has been to further develop that nomological stance. There are two main options open if one sets out to do so. One option is to abandon the call for something in the ontology that accounts for the temporal development of the entities that are posed in the primitive ontology. A central result of this paper is that this can be done in a clear and consistent manner. The mere fact of the existence of primitive ontology approaches to quantum mechanics and their empirical adequacy refutes the widespread impression that quantum mechanics contradicts Humeanism -- more precisely, Lewis' thesis of Humean supervenience: the world according to quantum mechanics can be ``a vast mosaic of local matters of particular fact'', namely the spatio-temporal distribution of the elements posed in the primitive ontology such as particle positions. The universal wave-function and the laws of quantum mechanics supervene on this distribution. They are nothing more than devices of economical bookkeeping, there being no real connections among the elements of the primitive ontology (such as a real relation of entanglement); there is only a non-dynamical background structure of space-time geometry into which these elements are inserted.
The other option is to conceive the elements of the primitive ontology as including a disposition to develop in a certain manner in time. That disposition then grounds the law of motion in the sense that the law expresses or reveals that disposition. Given the fact of quantum non-locality, that disposition has to be conceived as a holistic property of the elements of the primitive ontology -- in non-relativistic Bohmian mechanics, a property of all the particles at a given time taken together (in a relativistic setting, ``given time'' will have to be appropriately understood, for example by specifying a space-like hypersurface). One important message of our analysis is that the nomological interpretation can be consistently applied only to the fundamental wave-function referring to all local beables in the universe together. This fundamental universal level has to be distinguished from the effective description of subsystems, which is, in the end, what the usual quantum formalism is about.
The holistic disposition grounding the law is part and parcel of the primitive ontology, since it is not derived from anything and belongs to that what the quantum formalism represents or refers to. It is a beable existing in three-dimensional space, albeit not a local one. A further major result of this paper hence is that in endorsing dispositionalism, one can include something in the ontology that accounts for the temporal development of the local beables, without having to resort to accepting the wave-function in configuration space as an element of physical reality. One advantage of the particle ontology is that the disposition can be understood as a holistic property of the local beables themselves, whereas in the GRW-theory, spatial localization of the mass-density or flashes has to be regarded as the manifestation of the disposition (see Dorato and Esfeld 2010 for dispositions in GRW).
On the question of Humeanism versus dispositionalism we side with dispositionalism, since we take it to be a sound demand to call for something in the ontology that accounts for the temporal development of the elements of physical reality and that grounds the law of motion, thus providing for real connections in nature. Let us stress once again that the dualism of properties to which we are thus committed is only an appearance, a mere heuristic device to introduce dispositions. As explained above, there are not different types of properties assigned to the elements of the primitive ontology, spatial positions and velocities as categorical, local properties and in addition to that a holistic, dispositional property for a certain temporal development of these entities that fixes the velocities. There simply are beables that in being localized have the common disposition to develop in a certain manner in time. However, we are aware of the fact that clearing up the relationship between position and the disposition for temporal development (particles' positions and their holistic disposition of motion in Bohmian mechanics) is an open research project rather than a subject with established results. But the mere possibility of opening up that research question may be taken to be a further advantage of dispositionalism, showing that it is a promising programme for further research in the foundations of physics.\\

\begin{center} \textbf{Acknowledgements} \end{center}
We are grateful to two anonymous referees for helpful comments on the first version of the paper.\\

\newpage 
\section*{References}

\small{
 \noindent  Albert, David Z. (1996): ``Elementary quantum metaphysics''. In: J. T. Cushing, A. Fine and S. Goldstein (eds.): Bohmian mechanics and quantum theory: an appraisal. Dordrecht: Kluwer. Pp. 277-284.

\vspace{0.3cm}   \noindent  Allori, Valia, Goldstein, Sheldon, Tumulka, Roderich and Zangh\`i, Nino (2008): ``On the common structure of Bohmian mechanics and the Ghirardi-Rimini-Weber theory''. British Journal for the Philosophy of Science 59, pp. 353-389.

\vspace{0.3cm}   \noindent  Beebee, Helen (2006): ``Does anything hold the universe together?''. Synthese 149, pp. 509-533. 
  
\vspace{0.3cm}   \noindent  Bell, John S. (1987): Speakable and unspeakable in quantum mechanics. Cambridge: Cambridge University Press. 

\vspace{0.3cm}   \noindent  Belot, Gordon (2012): ``Quantum states for primitive ontologists. A case study''. European Journal for Philosophy of Science 2, pp. 67-83. 

\vspace{0.3cm}   \noindent  Bird, Alexander (2007): Nature's metaphysics. Laws and properties. Oxford: Oxford University Press. 

\vspace{0.3cm}   \noindent  Bohm, David (1952): ``A suggested interpretation of the quantum theory in terms of `hidden' variables''. Physical Review 85, pp. 166-193. 

\vspace{0.3cm}   \noindent  Bohm, David and Hiley, Basil (1993): The undivided universe. An ontological interpretation of quantum theory. London: Routledge. 

\vspace{0.3cm}   \noindent  Callender, Craig (unpublished): ``Discussion: the redundancy argument against Bohm's theory''. \\
http://philosophyfaculty.ucsd.edu/faculty/ccallender/publications.shtml

\vspace{0.3cm}   \noindent  Callender, Craig and Weingard, Robert (1997): ``Problems for Bohm's theory''. The Monist 80, pp. 24-43.

\vspace{0.3cm}   \noindent  de Broglie, Louis (1928): ``La nouvelle dynamique des quanta''. Electrons et photons: Rapports et discussions du cinqui\`eme Conseil de physique tenu \`a Bruxelles du 24 au 29 octobre 1927 sous les auspices de l'Institut international de physique Solvay. Paris: Gauthier-Villars. Pp. 105-132. 

\vspace{0.3cm}   \noindent  Dorato, Mauro (2005): The software of the universe. An introduction to the history and philosophy of laws of nature. Aldershot: Ashgate. 

\vspace{0.3cm}   \noindent  Dorato, Mauro and Esfeld, Michael (2010): ``GRW as an ontology of dispositions''. Studies in History and Philosophy of Modern Physics 41, pp. 41-49. 

\vspace{0.3cm}   \noindent  D\"urr, Detlef, Goldstein, Sheldon and Zangh\`i, Nino (1992): ``Quantum equilibrium and the origin of absolute uncertainty''. Journal of Statistical Physics 67, pp. 843-907. Reprinted in D. D\"urr, S. Goldstein and N. Zangh\`i: Quantum physics without quantum philosophy. Berlin: Springer 2013. Ch.2.

\vspace{0.3cm}   \noindent  D\"urr, Detlef, Goldstein, Sheldon and Zangh\`i, Nino (1997): ``Bohmian mechanics and the meaning of the wave-function''. In: R. S. Cohen, M. Horne and J. Stachel (eds.): Experimental metaphysics. Quantum mechanical studies for Abner Shimony. Volume 1. Dordrecht: Kluwer. Pp. 25-38. 

\vspace{0.3cm}   \noindent  D\"urr, Detlef and Teufel, Stefan (2009): Bohmian mechanics. The physics and mathematics of quantum theory. Berlin: Springer.
Forrest, Peter (1988): Quantum metaphysics. Oxford: Blackwell.

\vspace{0.3cm}   \noindent  Ghirardi, Gian Carlo, Grassi, Renata and Benatti, Fabio (1995): ``Describing the macroscopic world: Closing the circle within the dynamical reduction program''. Foundations of Physics 25, pp. 5-38.

\vspace{0.3cm}   \noindent  Goldstein, Sheldon (1998): ``Quantum theory without observers''. Physics Today 51, March pp. 42-46 \& April pp. 38-42. 

\vspace{0.3cm}   \noindent  Goldstein, Sheldon and Teufel, Stefan (2001): ``Quantum spacetime without observers: Ontological clarity and the conceptual foundations of quantum gravity''. In C. Callender and N. Huggett (eds.): Physics meets philosophy at the Planck scale. Cambridge: Cambridge University Press. Pp. 275-289. Reprinted in D. D\"urr, S. Goldstein and N. Zangh\`i: Quantum physics without quantum philosophy. Berlin: Springer 2013. Chapter 11.

\vspace{0.3cm}   \noindent  Goldstein, Sheldon and Zangh\`i, Nino (2012): ``Reality and the role of the wavefunction''. In: D. D\"urr, S. Goldstein and N. Zangh\`i: Quantum physics without quantum philosophy. Berlin: Springer. Chapter 12.

\vspace{0.3cm}   \noindent  Holland, Peter R. (1993): The quantum theory of motion. An account of the de Broglie-Bohm causal interpretation of quantum mechanics. Cambridge: Cambridge University Press. 

\vspace{0.3cm}   \noindent  Hume, David (1748): An enquiry concerning human understanding. 

\vspace{0.3cm}   \noindent  Lewis, David (1973): Counterfactuals. Oxford: Blackwell. 

\vspace{0.3cm}   \noindent  Lewis, David (1986a): Philosophical papers. Volume 2. Oxford: Oxford University Press. 

\vspace{0.3cm}   \noindent  Lewis, David (1986b): On the Plurality of Worlds. Oxford: Blackwell. 

\vspace{0.3cm}   \noindent  Loewer, Barry (1996): ``Humean Supervenience''. Philosophical Topics 24, pp. 101-127. 

\vspace{0.3cm}   \noindent  Maudlin, Tim (2007): The metaphysics within physics. Oxford: Oxford University Press.

\vspace{0.3cm}   \noindent  Monton, Bradley (2002): ``Wave-function ontology''. Synthese 130, pp. 265-277. 

\vspace{0.3cm}   \noindent  Monton, Bradley (2006): ``Quantum mechanics and 3N-dimensional space''. Philosophy of Science 73, pp. 778-789.

\vspace{0.3cm}   \noindent  Mumford, Stephen (2004): Laws in nature. London: Routledge. 

\vspace{0.3cm}   \noindent  Norsen, Travis (2010): ``The theory of (exclusively) local beables''. Foundations of Physics 40, pp. 1858-1884.

\vspace{0.3cm}   \noindent  Pusey, Matthew F., Barrett, Jonathan and Rudolph, Terry (2012): ``On the reality of the quantum state''. Nature Physics 8, pp. 475-478. 

\vspace{0.3cm}   \noindent  Su\'arez, Mauricio (2009): ``The many metaphysics within physics: Essay Review of \emph{The metaphysics within physics} by Tim Maudlin''. Studies in History and Philosophy of Modern Physics 40, pp. 273-276. 

\vspace{0.3cm}   \noindent  Teller, Paul (1986): ``Relational holism and quantum mechanics''. British Journal for the Philosophy of Science 37, pp. 71-81.

\vspace{0.3cm}   \noindent  Thomson-Jones, Martin (2012): ``Dispositions and quantum mechanics''. Manuscript. http://www.oberlin.edu/faculty/mthomson-jones/research.html (downloaded 31 May 2012). 

\vspace{0.3cm}   \noindent  Tumulka, Roderich (2006): ``A relativistic version of the Ghirardi-Rimini-Weber model''. Journal of Statistical Physics 125, pp. 821-840.
}

\end{document}